\documentclass[12pt]{article}
\usepackage{amsmath} 
\usepackage{color}
\input{epsf} 
\def\ltap{\raisebox{-.4ex}{\rlap{$\,\sim\,$}} \raisebox{.4ex}{$\,<\,$}}

\newcommand\as{\alpha_{\mathrm{S}}}

\def\beq{\begin{equation}} 
\def\eeq{\end{equation}} 

\def\to{\rightarrow}

\def\b0{b_0}

\newcommand{\ccaption}[2]{
Â Â Â \begin{center}
Â Â Â \parbox{0.85\textwidth}{
Â Â Â Â Â \caption[#1]{\small{{#2}}}
Â Â Â Â Â }
Â Â Â \end{center}
Â Â Â }

\begin{document} 

\begin{titlepage}
\renewcommand{\thefootnote}{\fnsymbol{footnote}}
\begin{flushright}
Â Â Â Â \end{flushright}
\par \vspace{10mm}

\begin{center}
{\Large \bf
Higgs production through gluon fusion: \\[0.3cm]
updated cross sections at the Tevatron and the LHC}
\end{center}
\par \vspace{2mm}
\begin{center}
{\bf Daniel de Florian}${}^{(a,b)}$\footnote{deflo@df.uba.ar} and
{\bf Massimiliano Grazzini}${}^{(c)}$\footnote{grazzini@fi.infn.it}\\

\vspace{5mm}

${}^{(a)}$Departamento de F\'\i sica, FCEYN, Universidad de Buenos Aires, \\
(1428) Pabell\'on 1, Ciudad Universitaria, Capital Federal, Argentina\\

${}^{(b)}$Physics Department, Brookhaven National Laboratory, Upton, New York 11973, USA\\

${}^{(c)}$INFN, Sezione di Firenze and Dipartimento di Fisica, Universit\`a di Firenze,\\ I-50019 Sesto Fiorentino, Florence, Italy\\

\vspace{5mm}

\end{center}

\par \vspace{2mm}
\begin{center} {\large \bf Abstract} \end{center}
\begin{quote}
\pretolerance 10000

We present updated predictions for the total cross section for Higgs boson production by gluon--gluon fusion in hadron collisions.
Our calculation includes the most advanced theoretical information available at present for this observable: soft-gluon resummation up to next-to-next-to-leading logarithmic
accuracy, the exact treatment of the bottom-quark contribution up to next-to-leading order, and two-loop electroweak effects.
We adopt the most recent parametrization of parton distribution functions at next-to-next-to-leading order, and we evaluate the corresponding uncertainties.
In comparison with
our previous central predictions, at the Tevatron the difference
ranges from $+9\%$ for $m_H=115$ GeV to $-9\%$ for $m_H=200$ GeV.
At the LHC the cross section is instead significantly increased.
The effect goes from $+30\%$ for $m_H=115$ GeV to $+9\%$ for $m_H=300$ GeV,
and is mostly
due to
the new parton distribution functions.
We also provide new predictions for the LHC at $\sqrt{s}=10$ TeV.

\end{quote}

\vspace*{\fill}
\begin{flushleft}
January 2009
\end{flushleft}
\end{titlepage}

\setcounter{footnote}{1}
\renewcommand{\thefootnote}{\fnsymbol{footnote}}

The Higgs boson is a key ingredient of the Standard Model (SM),
but it has so far eluded experimental discovery. 
Direct searches at LEP lead to a 95 $\%$ CL lower limit of $m_H>114.4$ GeV on the mass $m_H$ of the SM Higgs boson \cite{Barate:2003sz}.
At the LHC the Higgs boson can be discovered over the full mass range up to $m_H\sim 1$ TeV within a few
years of running. At the Tevatron, the CDF and D0 experiments are now becoming sensitive to a Higgs signal at $m_H\sim 170$ GeV \cite{Qian:2008cm}.

The dominant mechanism for SM Higgs boson production at hadron colliders is gluon-gluon fusion,
through a heavy-quark (mainly, top-quark) loop. When combined with the
decay channels $H\to\gamma\gamma$ and $H\to Z Z$ , this production mechanism is one of the most important
for Higgs boson searches and studies over the entire range, $100\,{\rm GeV} \ltap m_H \ltap 1\, {\rm TeV}$, of Higgs boson
mass to be investigated at the LHC.
In the mass range $140\,{\rm GeV}\ltap m_H \ltap 180\, {\rm GeV}$,
gluon fusion, followed by the decay $H\to WW\to l\nu l\nu$,
offers the main discovery channel of the Higgs boson at the LHC and
also at the Tevatron,
thanks to the strong angular correlations of the charged leptons in the final state.

In QCD perturbation theory the leading order (LO) contribution
to the $gg\to H$ cross section is proportional to $\as^2$,
$\as$ being the QCD coupling.
The QCD corrections have been computed at next-to-leading order (NLO) \cite{Dawson:1991zj,Djouadi:1991tk}
in the heavy-top limit,
and with full dependence on
the masses of the top and bottom quarks \cite{Djouadi:1991tk}.
Next-to-next-to-leading order (NNLO)
corrections have been obtained in the heavy-top limit \cite{NNLOtotal}.
These QCD corrections, which are dominated
by radiation of soft and virtual gluons \cite{soft},
lead to a substantial increase of the LO result.
The QCD computation up to NNLO has been consistently
improved by adding the resummation of soft-gluon logarithmic contributions, up
to next-to-next-to-leading logarithmic (NNLL) accuracy  \cite{Catani:2003zt}.
The ensuing NNLL+NNLO results
are nicely confirmed by the more recent computation 
\cite{Moch:2005ky}-\cite{Idilbi:2005ni} of some of the soft-gluon terms at N$^3$LO\footnote{Note, however, that
the perturbative information available at present does not allow to consistently extend
the computation to N$^3$LL accuracy.}.

In this Letter we present an update to the NNLL+NNLO results of Ref.~\cite{Catani:2003zt}.
We refrain from repeating the theoretical details of the calculation, 
which can be found in Ref.~\cite{Catani:2003zt}, and we only describe the main changes implemented in the present work.
In our previous analysis the benchmark predictions were obtained using
the MRST2002 NNLO partons \cite{Martin:2002aw}. Since then, there
have been important modifications in
the extraction of parton distribution functions (PDFs). The
calculation of the NNLO splitting functions\footnote{The
MRST2002 fit was based on approximate expressions for the evolution kernels \cite{vanNeerven:2000wp}.}
was completed \cite{nnlokernels}.
Furthermore, a more sophisticated treatment of
heavy-quark thresholds has been introduced in \cite{Martin:2007bv},
resulting in a significant modification of the gluon and light quark densities in the 
recent MSTW2008 set \cite{Martin:2009iq}.
For example, at $x\sim 0.01$, relevant for a Higgs boson of $m_H\sim 120$ GeV at the LHC,
the gluon distribution increases by about $6\%$ with respect to the MRST2002 fit.
The value of $\as(m_Z)$ also had a non-negligible change from 0.1154 to
0.1171. Considering that the total cross section is completely dominated by the
gluon-gluon fusion channel, and that the lowest order contribution starts at
${\cal O}(\as^2)$, with sizeable corrections at higher orders, it is not
surprising that the mere change from
MRST2002 to MSTW2008 partons
can result in an increase of more
than $10\%$ in the production cross section, 
making an update mandatory.
The change in the PDFs does not result in such a dramatic
increase of the cross section
at the Tevatron,
since at $x\sim 0.06$, relevant for the production of a Higgs boson of $m_H\sim 120$ GeV,
the gluon distribution is reduced by about $4\%$, but the decrease is partially compensated by the
increase of the partonic cross section due to the larger coupling constant.

Besides the important effect of the PDFs, there are other
theoretical reasons for revisiting the computation of the Higgs cross section at hadron colliders.
In particular there has been an important effort to evaluate the electroweak (EW) corrections arising
from $W$ and $Z$ boson coupling to the Higgs and to both light and heavy quarks in the loop \cite{ew}.
The recent computation of Ref.~\cite{Actis:2008ug} takes into account those
contributions by avoiding the complications in the two-particle threshold using the complex-mass scheme.
The EW corrections turn out to be of the order
of a few percent, with a sign depending on the Higgs mass. The main uncertainty in the EW
analysis comes from the fact that it is not completely clear how to take them into account in practice.
In the {\it partial factorization} scheme of Ref.~\cite{Actis:2008ug}
the EW correction applies
only to the LO result. In the {\em complete factorization} scheme instead, the EW correction multiplies the full QCD corrected cross section.
Since QCD corrections are sizeable, the latter choice has a
non-negligible
effect on the actual impact of EW corrections in the computation.
The recent analysis of higher-order QCD and EW
corrections presented in
Ref.~\cite{Anastasiou:2008tj}, performed on the basis of an effective Lagrangian approach,
supports the complete factorization hypothesis, suggesting that
EW corrections become, to a good approximation, a multiplicative
factor of the full QCD expansion.

The predictions we present below are obtained as follows. We first consider the top-quark 
contribution in the loop, and perform the calculation up NNLL+NNLO
in the large-$m_t$ limit. 
The result is rescaled by the exact $m_t$ dependent Born cross section, since this is known 
to be an excellent approximation for the top-quark contribution. This resummed top-quark contribution 
provides the bulk of the Higgs cross section at hadron colliders.
We then consider the bottom-quark contribution (more precisely, the bottom contribution and 
the top-bottom interference). Since in this case the effective theory approach is not applicable,
we follow Ref. \cite{Anastasiou:2008tj} and we include this contribution up to NLO only
(but still computed with NNLO MSTW2008 partons), by using the program HIGLU \cite{Djouadi:1991tk}.
Finally, we correct the result by including the EW effects evaluated in Ref.~\cite{Actis:2008ug}
in the complete factorization hypothesis.
We set the heavy-quark masses to $m_t=170.9$ GeV and $m_b=4.75$ GeV, the latter consistently
with the MSTW2008 set.
Our central predictions ($\sigma^{\rm best}$) are obtained by setting the
factorization ($\mu_F$) and renormalization ($\mu_R$) scales equal to the Higgs boson mass.

Our results for the Tevatron at $\sqrt{s}=1.96$ TeV and the LHC at $\sqrt{s}=10$ TeV and $\sqrt{s}=14$ TeV are presented
in Tables 1, 2 and 3, respectively.
Comparing to our previous predictions (see Tables 1 and 2 of Ref.~\cite{Catani:2003zt}), the cross sections change significantly. At the Tevatron the effect
ranges from $+9\%$ for $m_H=115$ GeV to $-9\%$ for $m_H=200$ GeV.
At the LHC  the effect goes from $+30\%$ for $m_H=115$ GeV to $+9\%$ 
for $m_H=300$ GeV. It is worth noticing that at the LHC more than half of the increase arises
from the modification in the gluon distribution and the coupling constant.

The bottom contribution, dominated by bottom-top interference, is small and negative. 
The different treatment of this contribution with respect to the 
previous analysis \cite{Catani:2003zt} results in an increase of the cross 
section from about $7\%$ ($m_H=115$ GeV) to $4\%$ 
($m_H=200$ GeV) at the Tevatron and from $9\%$ ($m_H=110$ GeV) to $2\%$ ($m_H=300$ GeV) 
at the LHC. The inclusion of the EW corrections results in an increase of the cross
section by about $5\%$
for $m_H\ltap 160$ GeV,
and a decrease by about $2\%$ for 
$200\,{\rm GeV}\ltap m_H \ltap 300\, {\rm GeV}$.

{ Our results for the Tevatron can be compared to those recently presented in 
Ref. \cite{Anastasiou:2008tj}, obtained using the same set of PDFs. Besides the different
choice for the bottom-quark mass and the implementation of an effective Lagrangian calculation for
the EW corrections, the main difference with our work arises
in the calculation of the top-quark contribution to the cross section.
In  Ref.~\cite{Anastasiou:2008tj} the latter contribution is computed up to NNLO
but choosing $\mu_F=\mu_R=m_H/2$,
as an attempt to mimic the effects of soft-gluon resummation beyond NNLO.
The final numerical differences at the Tevatron turn out to be small and 
of the order of a few {\em per mille} at the lowest masses, increasing to $2.5\%$ at $m_H=200$ GeV. }

We now discuss the various sources of uncertainty affecting the 
cross sections presented in Tables 1, 2 and 3. The uncertainty basically has two origins:
the one coming from the partonic cross sections, and the one arising from our limited
knowledge of the PDFs.

Uncalculated higher-order QCD radiative corrections are the most important source of uncertainty
on the partonic cross section. A method, which is customarily used in perturbative QCD
calculations, to estimate their size is to vary the renormalization and factorization scales around
the hard scale $m_H$. In general, this procedure can only give a lower limit on the {\it true}
uncertainty.
The uncertainty is quantified here as in Ref.~\cite{Catani:2003zt}: we vary
independently $\mu_F$ and $\mu_R$ 
in the range $0.5 m_H\leq \mu_F,\mu_R\leq 2 m_H$, with the constraint
$0.5 \leq \mu_F/\mu_R \leq 2$. 
The results are reported in Tables 1,2 and 3.
The scale uncertainty is about $\pm 9-10\%$ at the Tevatron and ranges from about $\pm 10\%$ ($m_H=110$ GeV)
to about $\pm 7\%$ ($m_H=300$ GeV) at the LHC ($\sqrt{s}=10$ and 14 TeV).
These results are consistent with those of Ref.~\cite{Catani:2003zt};
in particular, we note that the effect of scale variations
in our resummed calculation
is considerably reduced
with respect to the corresponding NNLO result.
At NNLO the scale uncertainty is about $\pm 14\%$ at the Tevatron
and ranges from about $\pm 12\%$ ($m_H=110$ GeV)
to about $\pm 9\%$ ($m_H=300$ GeV) at the LHC ($\sqrt{s}=10$ and 14 TeV).
The reduction is more sizeable at the Tevatron, where the resummation effect
is more important.

Another source of perturbative uncertainty on the partonic cross sections comes from the implementation of
the EW corrections. Our results are obtained in the complete factorization scheme discussed above.
The partial factorization scheme would lead to a change of our results ranging
from about $-3 \%$ ($m_H=115$ GeV) to $+2\%$ ($m_H=200$ GeV) at the Tevatron
and from about $-3 \%$ ($m_H=110$ GeV) to $+1\%$ ($m_H=300$ GeV) at the LHC.

A different
source of perturbative uncertainty comes from the use of the
large-$m_t$ approximation in the computation of the partonic cross section beyond LO.
The comparison between
the exact NLO cross section and the one obtained in the large-$m_t$ approximation (but rescaled with
the full Born result, including its exact dependence on $m_t$) shows that the approximation
works well also for $m_H > m_t$. This is not accidental: the higher-order contributions to
the cross section are dominated by relatively soft radiation, which is weakly sensitive to the mass of the
heavy quark in the loop at Born level. This feature persists at NNLO and thus it is natural to assume that having normalized our resummed result with the exact $m_t$-dependent cross section, the uncertainty due to
the large-$m_t$ approximation should be of the order of few percent, as it is at NLO.
{ The effect of finite-$m_t$ corrections is discussed in Refs.~\cite{Schreck:2007um,Marzani:2008az}.}

The other important source of uncertainty in the cross section is
the one coming from PDFs.
Modern PDF
sets
let the user estimate the experimental uncertainty
originating from the accuracy of the data points used to perform the fit.
The most recent MSTW2008 NNLO set provides 40 different grids that allow us
to evaluate the experimental uncertainties according
to the procedure discussed in Ref.~\cite{Martin:2002aw}.
The outcoming uncertainties (at $90\%$ CL) are reported in Tables 1,2 and 3.
At the Tevatron the effect ranges from $\pm 6\%$ ($m_H=115$ GeV) to about
$\pm 10\%$ ($m_H=200$ GeV), while at the LHC it
is about $\pm 3\%$ ($\pm 3-4\%$ at $\sqrt{s}=10$ TeV)
in the mass range we have considered.
We note that at the LHC the PDF uncertainty is quite small,
and, in particular,
it is substantially smaller than the uncertainty from missing higher-order perturbative contributions,
as estimated from scale variations.
On the contrary, the PDF uncertainty at the Tevatron is larger. This is a consequence of the fact that
the Tevatron probes relatively larger values of $x$, where the gluon density
is less constrained.
By using the MRST2006 set \cite{Martin:2007bv} our result for $m_H=170$ GeV at the Tevatron would be
$\sigma^{\rm best (2006)}=0.395^{+0.036}_{-0.031}\,({\rm scale})^{+0.017}_{-0.015}\,({\rm PDF})$,
which is $13\%$ smaller than the one reported in Table 1,
$\sigma^{\rm best}=0.349^{+0.032}_{-0.027}\,({\rm scale})^{+0.028}_{-0.029}\,({\rm PDF})$,
and marginally compatible with it.

We finally point out that
a related and important uncertainty is the one coming from the value of the QCD coupling.
In modern PDF sets $\as(m_Z)$ is obtained together with the parton densities
through a global fit to the available data,
and thus there will be a correlation between the error on $\as(m_Z)$ and that on the gluon density.
Since the gluon fusion process starts at ${\cal O}(\as^2)$, it is easy to foresee that the uncertainty on $\as(m_Z)$
may have an important impact on the cross section.
Neglecting correlations with the gluon density,
a $3\%$ uncertainty on $\as(m_Z)$ would
lead to an effect of about $\pm 9-10\%$ on the production cross section at both the Tevatron and the LHC.

To summarize, we have presented updated predictions for the cross section
for Higgs boson production at the Tevatron and the LHC. The results are based
on the most advanced theoretical information available at present for this observable, including 
soft-gluon resummation up to NNLL accuracy and two-loop EW corrections.
In comparison with
the central predictions of Ref.~\cite{Catani:2003zt},
at the Tevatron the difference ranges from
$+9\%$ to $-9\%$ for $115\, {\rm GeV} \ltap m_H \ltap\, 200\, {\rm GeV}$.
At the LHC the effect goes from $+30\%$ to $+9\%$ for
$115\, {\rm GeV} \ltap m_H \ltap\, 300\, {\rm GeV}$, and is mostly due to the new PDFs.
We have then reviewed \cite{Catani:2003zt} the uncertainties that affect the Higgs production cross section, and we have shown that they are still relatively large,
especially at the Tevatron.
The above uncertainties should be taken into account in Higgs boson searches and studies at both the Tevatron and the LHC.

We thank Babis Anastasiou and Christian Sturm for useful discussions.
We are grateful to Stefano Catani for helpful discussions and comments on the manuscript.
The work of D.deF. 
was partially supported by  ANPCYT, UBA-CyT and CONICET.
D.deF. is grateful to the US Department of Energy (Contract No. DE-AC02-98CH10886) for providing
the facilities essential for the 
completion of his work.

{\renewcommand{\arraystretch}{1.8} 
\begin{table}
\begin{center}
\begin{tabular}{|c|c|c|c|}
\hline
$m_H$ & $\sigma^{\rm best}$ & Scale & PDF\\
\hline
$100$ & $1.861$ & $^{+0.192}_{-0.174}$ & $^{+0.094}_{-0.101}$\\
$105$ & $1.618$ & $^{+0.165}_{-0.149}$ & $^{+0.085}_{-0.091}$\\
$110$ & $1.413$ & $^{+0.142}_{-0.127}$ & $^{+0.077}_{-0.083}$\\
$115$ & $1.240$ & $^{+0.123}_{-0.110}$ & $^{+0.070}_{-0.075}$\\
$120$ & $1.093$ & $^{+0.107}_{-0.095}$ & $^{+0.065}_{-0.069}$\\
$125$ & $0.967$ & $^{+0.094}_{-0.083}$ & $^{+0.059}_{-0.063}$\\
$130$ & $0.858$ & $^{+0.082}_{-0.072}$ & $^{+0.054}_{-0.058}$\\
$135$ & $0.764$ & $^{+0.073}_{-0.063}$ & $^{+0.050}_{-0.053}$\\
$140$ & $0.682$ & $^{+0.065}_{-0.056}$ & $^{+0.046}_{-0.049}$\\
$145$ & $0.611$ & $^{+0.057}_{-0.049}$ & $^{+0.042}_{-0.045}$\\
$150$ & $0.548$ & $^{+0.051}_{-0.044}$ & $^{+0.039}_{-0.042}$\\
$155$ & $0.492$ & $^{+0.045}_{-0.039}$ & $^{+0.036}_{-0.038}$\\
$160$ & $0.439$ & $^{+0.040}_{-0.034}$ & $^{+0.033}_{-0.035}$\\
$165$ & $0.389$ & $^{+0.036}_{-0.030}$ & $^{+0.030}_{-0.032}$\\
$170$ & $0.349$ & $^{+0.032}_{-0.027}$ & $^{+0.028}_{-0.029}$\\
$175$ & $0.314$ & $^{+0.029}_{-0.024}$ & $^{+0.026}_{-0.027}$\\
$180$ & $0.283$ & $^{+0.026}_{-0.021}$ & $^{+0.024}_{-0.025}$\\
$185$ & $0.255$ & $^{+0.023}_{-0.019}$ & $^{+0.022}_{-0.023}$\\
$190$ & $0.231$ & $^{+0.021}_{-0.017}$ & $^{+0.020}_{-0.021}$\\
$195$ & $0.210$ & $^{+0.019}_{-0.015}$ & $^{+0.019}_{-0.020}$\\
$200$ & $0.192$ & $^{+0.017}_{-0.014}$ & $^{+0.018}_{-0.019}$\\

\hline
\end{tabular}
\ccaption{}{\label{tab:tev}{\em Cross sections (in pb) at the
Tevatron ($\mu_F=\mu_R=m_H$) with $\sqrt{s}=1.96$~TeV, 
using the MSTW2008 \cite{Martin:2009iq} parton densities.}}
\end{center}
\end{table}}

{\renewcommand{\arraystretch}{2.0} Â \renewcommand{\tabcolsep}{1mm}
\begin{table}
\begin{center}
\begin{minipage}{0.31\textwidth}
\begin{tabular}{|c|c|c|c}
\hline
$m_H$ & $\sigma^{\rm best}$ & Scale & PDF\\
\hline
$100$ & $ 44.12 $& $^{+4.24}_{-4.44}$ & $^{+1.07}_{-1.39}$\\
$110$ & $ 36.99 $& $^{+3.43}_{-3.60}$ & $^{+0.88}_{-1.14}$\\
$120$ & $ 31.48 $& $^{+2.83}_{-2.96}$ & $^{+0.75}_{-0.96}$\\
$130$ & $ 27.11 $& $^{+2.35}_{-2.48}$ & $^{+0.64}_{-0.82}$\\
$140$ & $ 23.58 $& $^{+1.98}_{-2.10}$ & $^{+0.56}_{-0.71}$\\
$150$ & $ 20.69 $& $^{+1.69}_{-1.80}$ & $^{+0.50}_{-0.62}$\\
$160$ & $ 18.07 $& $^{+1.44}_{-1.53}$ & $^{+0.44}_{-0.55}$\\
\hline
\end{tabular}
\end{minipage}
\begin{minipage}{0.31\textwidth}
\begin{tabular}{|c|c|c|c}
\hline
$m_H$ & $\sigma^{\rm best}$ & Scale & PDF\\
\hline
$170$ & $ 15.63 $& $^{+1.22}_{-1.30}$ & $^{+0.39}_{-0.48}$\\
$180$ & $ 13.78 $& $^{+1.05}_{-1.12}$ & $^{+0.35}_{-0.42}$\\
$190$ & $ 12.20 $& $^{+0.91}_{-0.97}$ & $^{+0.32}_{-0.38}$\\
$200$ & $ 10.97 $& $^{+0.80}_{-0.86}$ & $^{+0.29}_{-0.35}$\\
$210$ & $ 9.98 $& $^{+0.72}_{-0.77}$ & $^{+0.27}_{-0.32}$\\
$220$ & $ 9.14 $& $^{+0.64}_{-0.69}$ & $^{+0.26}_{-0.30}$\\
$230$ & $ 8.42 $& $^{+0.58}_{-0.63}$ & $^{+0.24}_{-0.28}$\\
\hline
\end{tabular}
\end{minipage}
\begin{minipage}{0.31\textwidth}
\begin{tabular}{|c|c|c|c}
\hline
$m_H$ & $\sigma^{\rm best}$ & Scale & PDF\\
\hline
$240$ & $ 7.81 $& $^{+0.53}_{-0.58}$ & $^{+0.23}_{-0.26}$\\
$250$ & $ 7.29 $& $^{+0.49}_{-0.53}$ & $^{+0.22}_{-0.25}$\\
$260$ & $ 6.83 $& $^{+0.45}_{-0.49}$ & $^{+0.21}_{-0.24}$\\
$270$ & $ 6.44 $& $^{+0.42}_{-0.46}$ & $^{+0.21}_{-0.23}$\\
$280$ & $ 6.11 $& $^{+0.40}_{-0.43}$ & $^{+0.20}_{-0.22}$\\
$290$ & $ 5.83 $& $^{+0.37}_{-0.40}$ & $^{+0.20}_{-0.22}$\\
$300$ & $ 5.61 $& $^{+0.37}_{-0.38}$ & $^{+0.19}_{-0.21}$\\
\hline
\end{tabular}
\end{minipage}
\ccaption{}{\label{tab:lhc10}{\em Cross sections (in pb)
at the LHC ($\mu_F=\mu_R=m_H$) with $\sqrt{s}=10$ TeV 
using the MSTW2008 \cite{Martin:2009iq} parton densities.}}
\end{center}
\end{table}}

{\renewcommand{\arraystretch}{2.0} Â \renewcommand{\tabcolsep}{1mm}
\begin{table}
\begin{center}
\begin{minipage}{0.31\textwidth}
\begin{tabular}{|c|c|c|c}
\hline
$m_H$ & $\sigma^{\rm best}$ & Scale & PDF\\
\hline
$100$ & $ 74.58 $& $^{+7.18}_{-7.54}$ & $^{+1.86}_{-2.45}$\\
$110$ & $ 63.29 $& $^{+5.87}_{-6.20}$ & $^{+1.54}_{-2.02}$\\
$120$ & $ 54.48 $& $^{+4.88}_{-5.18}$ & $^{+1.30}_{-1.70}$\\
$130$ & $ 47.44 $& $^{+4.12}_{-4.38}$ & $^{+1.12}_{-1.45}$\\
$140$ & $ 41.70 $& $^{+3.47}_{-3.75}$ & $^{+0.97}_{-1.25}$\\
$150$ & $ 36.95 $& $^{+3.02}_{-3.24}$ & $^{+0.85}_{-1.10}$\\
$160$ & $ 32.59 $& $^{+2.60}_{-2.79}$ & $^{+0.73}_{-0.97}$\\
\hline
\end{tabular}
\end{minipage}
\begin{minipage}{0.31\textwidth}
\begin{tabular}{|c|c|c|c}
\hline
$m_H$ & $\sigma^{\rm best}$ & Scale & PDF\\
\hline
$170$ & $ 28.46 $& $^{+2.22}_{-2.39}$ & $^{+0.65}_{-0.84}$\\
$180$ & $ 25.32 $& $^{+1.92}_{-2.08}$ & $^{+0.58}_{-0.74}$\\
$190$ & $ 22.63 $& $^{+1.68}_{-1.83}$ & $^{+0.52}_{-0.66}$\\
$200$ & $ 20.52 $& $^{+1.49}_{-1.63}$ & $^{+0.48}_{-0.60}$\\
$210$ & $ 18.82 $& $^{+1.34}_{-1.47}$ & $^{+0.45}_{-0.55}$\\
$220$ & $ 17.38 $& $^{+1.22}_{-1.33}$ & $^{+0.42}_{-0.51}$\\
$230$ & $ 16.15 $& $^{+1.11}_{-1.22}$ & $^{+0.39}_{-0.48}$\\
\hline
\end{tabular}
\end{minipage}
\begin{minipage}{0.31\textwidth}
\begin{tabular}{|c|c|c|c}
\hline
$m_H$ & $\sigma^{\rm best}$ & Scale & PDF\\
\hline
$240$ & $ 15.10 $& $^{+1.03}_{-1.12}$ & $^{+0.37}_{-0.45}$\\
$250$ & $ 14.19 $& $^{+0.95}_{-1.04}$ & $^{+0.36}_{-0.43}$\\
$260$ & $ 13.41 $& $^{+0.88}_{-0.97}$ & $^{+0.35}_{-0.41}$\\
$270$ & $ 12.74 $& $^{+0.83}_{-0.91}$ & $^{+0.33}_{-0.39}$\\
$280$ & $ 12.17 $& $^{+0.78}_{-0.86}$ & $^{+0.33}_{-0.38}$\\
$290$ & $ 11.71 $& $^{+0.74}_{-0.82}$ & $^{+0.32}_{-0.37}$\\
$300$ & $ 11.34 $& $^{+0.71}_{-0.78}$ & $^{+0.32}_{-0.36}$\\
\hline
\end{tabular}
\end{minipage}
\ccaption{}{\label{tab:lhc14}{\em Cross sections (in pb)
at the LHC ($\mu_F=\mu_R=m_H$) with $\sqrt{s}=14$ TeV 
using the MSTW2008 \cite{Martin:2009iq} parton densities.}}
\end{center}
\end{table}}

\end{document}